\documentclass{aa}
\usepackage{psfig}
\usepackage{natbib}
\usepackage{aas_macros}

\bibpunct{(}{)}{;}{a}{}{,}
\defcitealias{nyp+00}{Paper~I}

\begin{document}

\title{Population synthesis for double white dwarfs\\ 
II. Semi-detached systems: AM CVn stars}

\author{G. Nelemans\inst{1}, 
        S. F. Portegies Zwart\inst{2}\thanks{Hubble Fellow}, 
        F. Verbunt\inst{3} and 
        L. R. Yungelson\inst{1,4}}
\offprints{G. Nelemans, GijsN@astro.uva.nl}

\institute{Astronomical Institute ``Anton Pannekoek'', 
        Kruislaan 403, NL-1098 SJ Amsterdam, the Netherlands 
             \and
        Massachusetts Institute of Technology, Massachusetts Ave. 77,
                Cambridge MA 02139, USA
                \and 
           Astronomical Institute, Utrecht University,
           P.O.Box 80000, NL-3508 TA Utrecht, the Netherlands
             \and
             Institute of Astronomy of the Russian Academy of
             Sciences, 48 Pyatnitskaya Str., 109017 Moscow, Russia 
}

\date{received \today}

\titlerunning{Population synthesis for AM CVn stars}
\authorrunning{Nelemans, Portegies Zwart, Verbunt \& Yungelson}

\abstract{We study two models for AM CVn stars: white dwarfs accreting
  (i) from a helium white dwarf companion and (ii) from a helium-star
  donor. We show that in the first model possibly no accretion disk
  forms at the onset of mass transfer.  The stability and the rate of
  mass transfer then depend on the tidal coupling between the accretor
  and the orbital motion.  In the second model the formation of AM CVn
  stars may be prevented by detonation of the CO white dwarf accretor
  and the disruption of the system.  With the most favourable
  conditions for the formation of AM CVn stars we find a current
  Galactic birth rate of $6.8 \times 10^{-3}\,{\rm yr^{-1}}$.
  Unfavourable conditions give $1.1 \times 10^{-3}\,{\rm yr^{-1}}$.
  The expected total number of the systems in the Galaxy is $9.4\times
  10^{7}$ and $1.6 \times 10^{7}$, respectively. We model very simple
  selection effects to get some idea about the currently expected
  observable population and discuss the (quite good) agreement with
  the observed systems.
\keywords{stars: white dwarfs -- stars: statistics -- binaries: close
 -- binaries: evolution}
}

\maketitle

\section{Introduction}

AM CVn stars are helium-rich faint blue objects that exhibit
variability on time-scales of $\sim$1000 seconds. \citet{sma67}
discovered that the prototype of the class, AM CVn (=HZ 29), shows
photometric variability with a period $\sim 18$ min and suggested that
it is a binary. Immediately \citet{pac67} realised that it could be a
semi-detached pair of degenerate dwarfs in which mass transfer is
driven by loss of angular momentum due to gravitational wave
radiation. After flickering, typical for cataclysmic binary systems,
was found in AM CVn by \citet{wr72}, this model was used to explain
this system by \citet{ffw72}. There are currently 8 AM CVn candidates
(Table~\ref{obs}) which have been studied photometrically in great
detail (for reviews see \citet{ull94}, \citet{war95} and
\citet{sol95}).  AM CVn stars also attracted attention as possible
sources of gravitational waves \citep[and references therein]{hb00}.

After the introduction of the concept of common envelope evolution for
the formation of cataclysmic variables and X-ray binaries
\citep{pac76}, the formation of close double white dwarfs through two
of such phases was anticipated by \citet{ty79a,ty81}.  The emission of
gravitational waves would subsequently bring the two white dwarfs into
a semi-detached phase. \citet{nrs81} independently suggested this
scenario for the formation of AM CVn itself.  In an alternative
scenario the white dwarf donor is replaced by a helium star that
becomes semi-degenerate during the mass transfer \citep{it91}.

Throughout this paper we use the term AM CVn for binaries in which a
white dwarf accretes from an another white dwarf or from a
semi-degenerate helium star, irrespective how they would be classified
observationally.  

This paper continues our study on the formation and evolution of the
Galactic population of close double white dwarfs \citep{nvy+00,
  nyp+00}. Here we study the population that becomes semi-detached and
transfers mass in a stable way. In addition we examine the alternative
case where the donor is a semi-degenerate helium star.
  
In Sect.~\ref{model} we outline the evolution of binaries driven by
gravitational wave radiation. We review the models for the formation
of AM CVn stars and discuss the stability of the mass transfer in
Sect.\ref{formation}. The results of our population synthesis and a
comparison with observations are presented in
Sect.~\ref{sec:popsynth}.  The differences with previous studies are
discussed in Sect.~\ref{discussion} after which the conclusions
follow.

\section{Mass transfer in close binaries driven by gravitational wave
         radiation} 
\label{model}

The rate of angular momentum loss ($\dot{J}$) of a binary system
with a circular orbit due to gravitational wave radiation (GWR) is
\citep{ll71}:
\begin{equation}
\left ( \frac {\dot{J}}{J} \right )_{\rm GWR}  = -\frac{32}{5} \, \frac{G^3}{c^5} \frac{\,M \,m \,(M+m)}{a^{4}}.
\end{equation}
Here $M$ and $m$ are the masses of the two components and $a$ is their
orbital separation.

In a binary with stable mass transfer the change of the radius of the
donor exactly matches the change of its Roche lobe.  This condition
combined with an approximate equation for the size of the Roche lobe
\citep{pac67},
\begin{equation}\label{eq:r_L}
R_{\rm L } \approx 0.46 \, a \left(\frac{m}{ M + m}\right)^{1/3} \qquad \mbox{
  for } m < 0.8 M,
\end{equation}
may be used to derive the rate of mass transfer for a semi-detached
binary in which the mass transfer is driven by 
GWR \citep{pac67}
\begin{equation}\label{eq:mdot} 
\frac{\dot{m}}{m} = \left (
  \frac{\dot{J}}{J} \right )_{\rm GWR} \times \left [\frac{\zeta (m)}{2} +
\frac{5}{6} - \frac{m}{M} \right]^{-1}.
\end{equation}
Here $\zeta (m)$ is the logarithmic derivative of the radius of the
donor with respect to its mass ($\zeta \equiv$ d $\ln r/$ d $\ln m$).
For the mass transfer to be stable, the term in brackets must be
positive, i.e.
\begin{equation}\label{eq:q_II}
 q \equiv \frac{m}{M} < \frac{5}{6} + \frac{\zeta(m)}{2}.
\end{equation}
The mass transfer becomes dynamically unstable when this criterion is
violated, probably causing the binary components to coalesce
\citep{pw75,ty79a}.

\section{The nature of the mass donor: two formation
         scenarios}\label{formation} 

\subsection{Close double white dwarfs as AM CVn progenitors}
\label{formation:dwd}

From the spectra of AM CVn stars it is inferred that the transferred
material consists mainly of helium.  \citet{ffw72} suggested two
possibilities for the helium rich donor in AM CVn: (i) a zero-age
helium star with a mass of 0.4 - 0.5 \mbox{${\rm M}_{\sun}$}\, and (ii) a low-mass
degenerate helium white dwarf. The first possibility can be excluded
because the helium star would dominate the spectrum and cause the
accretor to have large radial velocity variations, none of which is
observed. Thus they concluded that AM CVn stars are interacting double
white dwarfs. Their \emph{direct progenitors} may be detached close
double white dwarfs which are brought into contact by loss of angular
momentum due to GWR within the lifetime of the Galactic disk (for
which we take 10 Gyr). The less massive white dwarf fills its Roche
lobe first and an AM CVn star is born, if the stars do not merge (see
Sect.~\ref{stability}). We discussed the formation of such double
white dwarfs in \citet{nyp+00}.

To calculate the stability of the mass transfer and the evolution of
the AM CVn system, one needs to know the mass-radius relation for white
dwarfs.  This depends on the temperature, chemical composition,
thickness of the envelope etc. of the white dwarf.  However,
\citet{pab00} have shown that after cooling for several 100\,Myr the
mass - radius relation for low-mass helium white dwarfs approaches the
relation for zero-temperature spheres. As most white dwarfs that may
form AM CVn stars are at least several 100\,Myr old at the moment of
contact \citep{ty96}, we apply the mass - radius relation for cold
spheres derived by \citet{zs69}, as corrected by \citet{rj84}.  For
helium white dwarfs with masses between 0.002 and 0.45 \mbox{${\rm M}_{\sun}$}, it can
be approximated to within 3\% by (in solar units)
\begin{equation}
\label{eq:deg}
R_{\rm ZS} \approx 0.0106 - 0.0064 \; \ln M_{\rm WD} + 0.0015 M_{\rm WD}^2.
\end{equation}
We apply the same equation for the radii of CO white dwarfs, since
the dependence on chemical composition is negligible in the range of
interest.

\subsection{Stability of the mass transfer between white dwarfs}
\label{stability}

\begin{figure}
\psfig{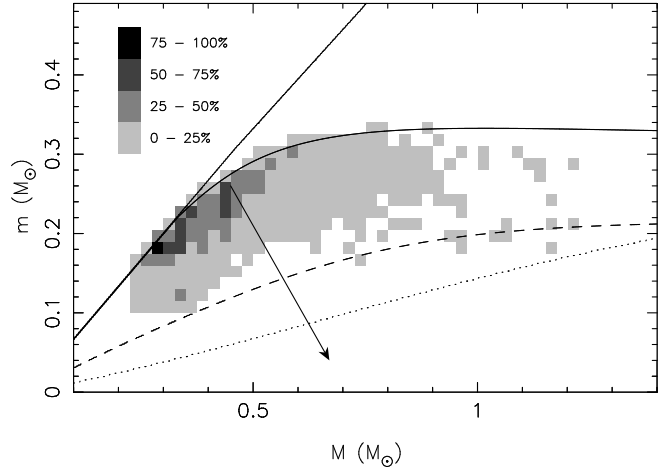}
\caption[]{
  Stability limits for mass transfer in close double white dwarfs.
  Above the upper solid curve the mass transfer is dynamically
  unstable. Below the lower solid line the systems have mass transfer
  rates below the Eddington limit in the case of an efficient tidal
  coupling between the accretor and the orbital motion. If the
  coupling is inefficient, this limit shifts down to the dashed line.
  Above the dotted line the stream hits the companion directly at the
  onset of the mass transfer and no accretion disk forms.  As the
  evolution proceeds (parallel to the arrow) a disk
  eventually forms. The gray shades give the current model birthrate
  distribution of AM CVn stars that form from close binary white
  dwarfs.  The shading is scaled as a fraction of the maximum
  birthrate per bin, which is $1.4 \times 10^{-4}$
    yr$^{-1}$.}
\label{fig:MsMp}
\end{figure}

In Fig.~\ref{fig:MsMp} we show the limiting mass ratio for
dynamically stable mass transfer (Eq.~(\ref{eq:q_II})) as the upper
solid line, with $\zeta (m)$ derived from Eq.~(\ref{eq:deg}).  The
initial mass-transfer rates, as given by Eq.~(\ref{eq:mdot}), can be
higher than the Eddington limit of the accretor \citep{ty79a}.  The
matter that cannot be accreted is lost from the system, taking along
some angular momentum. The binary system may remain stable even though
it loses extra angular momentum. However heating of the transferred
material, may cause it to expand and form a common envelope in which
the two white dwarfs most likely merge \citep{hw99}.  Therefore, we
impose the additional restriction to have the initial mass transfer
rate lower than the Eddington accretion limit for the companion
($\sim\!10^{-5}\,\mbox{${\rm M}_{\sun}$}$ yr$^{-1}$). This changes the limiting mass
ratio below which AM CVn stars can be formed to the lower solid line
in Fig.~\ref{fig:MsMp}. In this Figure we over-plotted our model
distribution of the current birthrate of AM CVn stars that form from
close binary white dwarfs (see Sect.~\ref{sec:popsynth}).

In the derivation of the Eq.~(\ref{eq:mdot}) it is implicitly assumed
that the secondary rotates synchronously with the orbital revolution
and that the angular momentum which is drained from the secondary is
restored to the orbital motion via tidal interaction between the
accretion disk and the donor star \citep[see, e.g.][and references
therein]{vr88}.

However, the orbital separation when Roche lobe overflow starts is
only about 0.1 $R_{\sun}$ and the formation of the accretion disk is
not obvious; the matter that leaves the vicinity of the first
Lagrangian point initially follows a ballistic trajectory, passing the
accreting star at a minimum distance of $\sim 10\%$ of the binary
separation \citep{ls75}, i.e. at a distance comparable to the radius
of a white dwarf $(\sim 0.01 R_{\sun})$.  So, the accretion stream may
well hit the surface of the accretor directly instead of forming an
accretion disk around it \citep{web84}.

The minimum distance at which the accretion stream passes the accretor
is computed by \citet{ls75} (their $\widetilde{\omega}_{\rm min}$),
which we fit with
\begin{eqnarray}\label{eq:rmin}
        \frac{r_{\rm min}}{a} & \approx & 0.04948 \; - \; 0.03815 \;
                                \log (q) \\ \nonumber 
                      & & + \; 0.04752 \; \log^2 (q) \; - \; 0.006973
                            \log^3 (q).
\end{eqnarray}
The value of $q$ at which the radius of the accretor 
equals $r_{\rm min}$ is presented as a dotted line in 
Fig.~\ref{fig:MsMp}; above this line the accretion stream hits the
white dwarfs' surface directly and no accretion disk is formed.

In absence of the disk the angular momentum of the stream is converted
into spin of the accretor and mechanisms other than disk - orbit
interaction are required to transport the angular momentum of the
donor back to the orbit. The small separation between the two stars
may result in tidal coupling between the accretor and the donor which
is in synchronous rotation with the orbital period.  The efficiency of
this process is uncertain \citep{sb76, cam84}, but if tidal coupling
between accretor and donor is efficient the stability limit for mass
transfer is the same as in the presence of the disk
(Eq.~(\ref{eq:q_II})). In the most extreme case all the angular
momentum carried with the accretion stream is lost from the binary
system.  The lost angular momentum can be approximated by the angular
momentum of the ring that would be formed in the case of a point-mass
accretor: $\dot{J}_{\dot{m}} = \dot{m} \sqrt{G M a r_{\rm h}}$, where
$r_{\rm h}$ is the radius of the ring in units of $a$.  This sink of
angular momentum leads to an additional term $-\sqrt{(1+q) r_{\rm h}}$
in the brackets in Eq.~(\ref{eq:mdot}).  As a result the condition 
for dynamically stable mass transfer 
becomes more rigorous:
\begin{equation}\label{eq:q_I}
 q < \frac{5}{6} + \frac{\zeta(m)}{2} -\sqrt{(1+q) r_{\rm h}}.
\end{equation}
This limit (with $r_{\rm h}$ given by \citet{vr88} and again the
additional restriction of a mass transfer rate below the Eddington
limit) is shown in Fig.~\ref{fig:MsMp} as the dashed line.

Figure~\ref{fig:MsMp} shows that, with our assumptions, none of the AM
CVn binaries which descend from double white dwarfs (which we will
call the \emph{white dwarf family}) forms a disk at the onset of mass
transfer. After about 10$^7$ yr, when the donor mass has decreased
below 0.05 \mbox{${\rm M}_{\sun}$} (see Fig.~\ref{fig:PMdot}) and the orbit has become
wider a disk will form.

Only one of the currently known 14 close double white dwarfs possibly
is an AM CVn progenitor; WD\,1704+481A has $P_{\rm orb} = 3.48$\,hr, $m
= 0.39 \pm 0.05\,\mbox{${\rm M}_{\sun}$}$ and $M = 0.56 \pm 0.05\,\mbox{${\rm M}_{\sun}$}$\ 
\citep{mmm+00}. It is close to the limit for dynamical stability, but
because the initial mass transfer rate is expected to be
super-Eddington it may merge.

\subsection{Binaries with low mass helium stars as AM CVn progenitors: a semi-degenerate mass donor}\label{formation:hewd}
          
Another way to form a helium transferring binary in the right period
range was first outlined by \citet{skh86}. They envisioned a neutron
star accretor, but the scenario for an AM CVn star, with a white dwarf
accretor, is essentially the same \citep{it91}. One starts with a
low-mass, non-degenerate helium burning star, a remnant of so-called
case B mass transfer, with a white dwarf companion.  If the components
are close enough, loss of angular momentum via GWR may result in Roche
lobe overflow before helium exhaustion in the stellar core.  Mass
transfer is stable if the ratio of the mass of the helium star (donor)
to the white dwarf (accretor) is smaller than $\sim1.2$
\citep{tf89,ef90}\footnote{This stability limit is the same as for
  hydrogen-rich stars with radiative envelopes.}.  When the mass of
the helium star decreases below $\sim\!0.2\,\mbox{${\rm M}_{\sun}$}$, core helium
burning stops and the star becomes semi-degenerate.  This causes the
exponent in the mass-radius relation to become negative and, as a
consequence, mass transfer causes the orbital period to increase.  The
minimum period is $\sim$10\,min.  With strongly decreasing mass
transfer rate the donor mass drops below 0.01\mbox{${\rm M}_{\sun}$}\ in a few Gyr,
while the period of the system increases up to $\sim\!1$\,hr; in the
right range to explain the AM~CVn stars. The luminosity of the donor
drops below $10^{-4}\,\mbox{${\rm R}_{\sun}$}$ and its effective temperature to several
thousand K.  We will call the AM CVn stars that formed in this way the
\emph{helium star family}. Note that in this scenario a disk
  will always form because the orbit is rather wide at the onset of
  the mass transfer. The equations for efficient coupling thus hold.

The progenitors of these helium stars have masses in the range 2.3 --
5 \mbox{${\rm M}_{\sun}$}. The importance of this scenario is enhanced by the long
lifetimes of the helium stars: $t_{\rm He} \approx 10^{7.15}
M^{-3.7}_{\rm He}$\,yr\ \citep{it85}, comparable to the main-sequence
lifetime of their progenitors, so that there is enough time to lose
angular momentum by gravitational wave radiation and start mass
transfer before the helium burning stops.

\begin{figure}
\centerline{\psfig{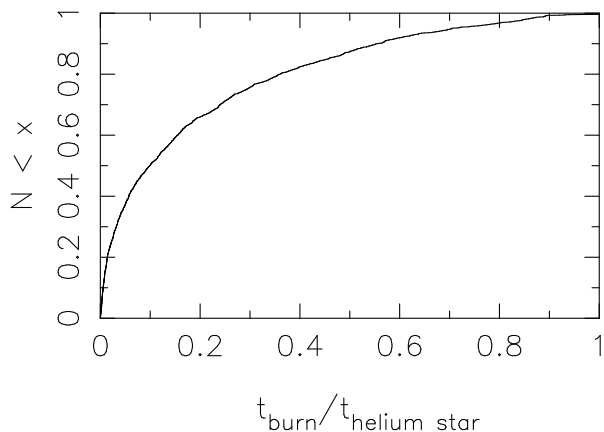}}
\caption[]{
  Cumulative distribution of the ratios of the helium burning time
  that occurred \emph{before} the mass transfer started and the total
  helium burning time for the model systems.
}
\label{fig:t_burn}
\end{figure}

Our simulation of the population of helium stars with white dwarf
companions, suggests that at the moment they get into contact the
majority of the helium stars are at the very beginning of core helium
burning. This is illustrated in Fig.~\ref{fig:t_burn}. Having this in
mind, we approximate the mass-radius relation for semi-degenerate
stars by a power-law fit to the results of computations of
\citet{tf89} for a 0.5\,\mbox{${\rm M}_{\sun}$}\ star, which filled its Roche lobe
shortly after the beginning of core helium burning (their model 1.1).
For the semi-degenerate part of the track we obtain (in solar
units):
\begin{equation}
\label{eq:semi}
R_{\rm TF} \approx 0.043 \; m^{-0.062}.
\end{equation}   
Trial computations with the relation $R \approx 0.029 \,
m^{-0.19}$\ from the model of \citet{skh86} which had $Y_{\rm c} =
0.26$\ at the onset of the mass transfer reveals a rather weak
dependence of our results on the mass - radius relation.

As noticed by \citet{skh86}, severe mass loss in the phase before the
period minimum, increases the thermal time-scale of the donor beyond
the age of the Galactic disk and thus prevents the donor from
becoming fully degenerate and keeps it semi-degenerate.

Another effect of the severe mass loss is the quenching of the helium
burning in the core. \citet{tf89} show that during the mass transfer
the central helium content hardly changes (especially for low mass
helium stars). Therefore, despite the formation of an outer convective
zone, which penetrates inward to regions where helium burning took
place, one would expect that in the majority of the systems the
transferred material is helium-rich down to very low donor masses.
However, donors with He-exhausted cores at the onset of mass transfer
($Y_{\rm c} \la 0.1$) may in the course of their evolution start to
transfer matter consisting of a carbon-oxygen mixture \citep[see Fig.~3
in][who used the same evolutionary code as Tutukov \& Fedorova]{ef90}.

\begin{figure}
\centerline{\psfig{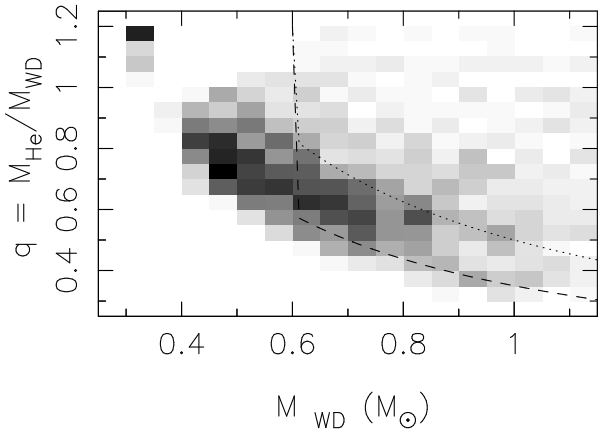}}
\caption[]{
  Distribution of helium stars with white dwarf companions that
  currently start stable mass transfer.  In the systems to the right
  of the dotted and dashed lines the white dwarfs accrete at least
  0.3\,\mbox{${\rm M}_{\sun}$} and 0.15\,\mbox{${\rm M}_{\sun}$}\ before ELD, respectively, at
  high accretion rates. These lines show two choices of mass limits
  above which, with our assumption, the binaries are disrupted by
  edge-lit detonation before they become AM CVn systems. The systems
  in the top left corner are binaries with helium white dwarf
  accretors. }
\label{hewd}
\end{figure}

Figure~\ref{hewd} shows the population of low-mass helium stars with
white dwarf companions which currently start mass transfer, (i.e. have
$q\la 1.2$), derived by means of population synthesis. It is possible
that only a fraction of them evolves into AM CVn stars. Upon Roche
lobe overflow, before the period minimum, most helium donors lose mass
at an almost constant rate close to $3 \,\times \,10^{-8}\,\mbox{${\rm M}_{\sun}$}$
yr$^{-1}$. Accretion of He at such rates by a carbon-oxygen (CO) white
dwarf may trigger a detonation in the layer of the accumulated matter
\citep{taa80a}. This may further cause the detonation of the
underlying CO dwarf, so-called edge-lit detonation, ELD \citep{liv90,
  lg91, ww94, la95}. The conditions for ELD to occur: the mass of
the white dwarf, the range of accretion rates, the mass of the
accumulated layer, etc.  are still actively debated.

However, examples computed by \citet{lt91} and \citet{ww94} show that
if $\dot{M} \ga 10^{-8}\,\mbox{${\rm M}_{\sun}$}$ yr$^{-1}$ the helium layer inevitably
detonates if $\Delta {\rm M_{\rm He}} \ga 0.3\,\mbox{${\rm M}_{\sun}$}$ and $M_{\rm CO}
\ga 0.6\,\mbox{${\rm M}_{\sun}$}$.  Therefore, as one of the extreme cases, we reject
all systems which satisfy these limits from the sample of progenitors
of AM CVn stars, assuming that they will be disrupted before the
helium star enters the semi-degenerate stage.  As another extreme,
we assume that only 0.15\,\mbox{${\rm M}_{\sun}$}\ has to be accreted for ELD (as a
compromise between results of \citet{lt91} and \citet{ww94}).  The
relevant cut-offs are shown in Fig.~\ref{hewd}.

For CO white dwarf accretors less massive than 0.6\,\mbox{${\rm M}_{\sun}$}\ we assume that
accretion of helium results in ``flashes'' in which the He layer is
ejected or lost via a common envelope formed due to expansion of the
layer. Such events may be repetitive.

\citet{lt91} show that for accretion rates below $10^{-8}\,\mbox{${\rm M}_{\sun}$}$
yr$^{-1}$ more than $\sim 0.4\,\mbox{${\rm M}_{\sun}$}$ helium has to be accreted before
detonation. For systems in which the donor becomes semi-degenerate and
the mass accretion rates are low we limit the accumulation of He only
by adopting the Chandrasekhar mass as a maximum to the total mass of
the accreting white dwarf.

\subsection{Summary: two extreme models for AM CVn progenitors}\label{sec:two_fam}  
          
We recognise two possibilities for each family of potential AM CVn
systems, which are: efficient or non-efficient tidal coupling between
the accretor and the orbital motion in the white dwarf family, and two
limits for the disruption of the accretors by ELD in the helium star
family.  We compute the populations for every possible solution and
combine them into two models: model I, in which there is no tidal
coupling and ELD is efficient in the destruction of potential
progenitor systems (an ``inefficient'' scenario for forming AM CVn
systems) and an ``efficient'' model II, in which there is an effective
tidal coupling and ELD is efficient only in systems with the most
massive donors.  However, we give the birth rates and number of the
objects for the four different solutions separately in
Table~\ref{tab:birthrates}.

\begin{figure*}
  \centerline{\psfig{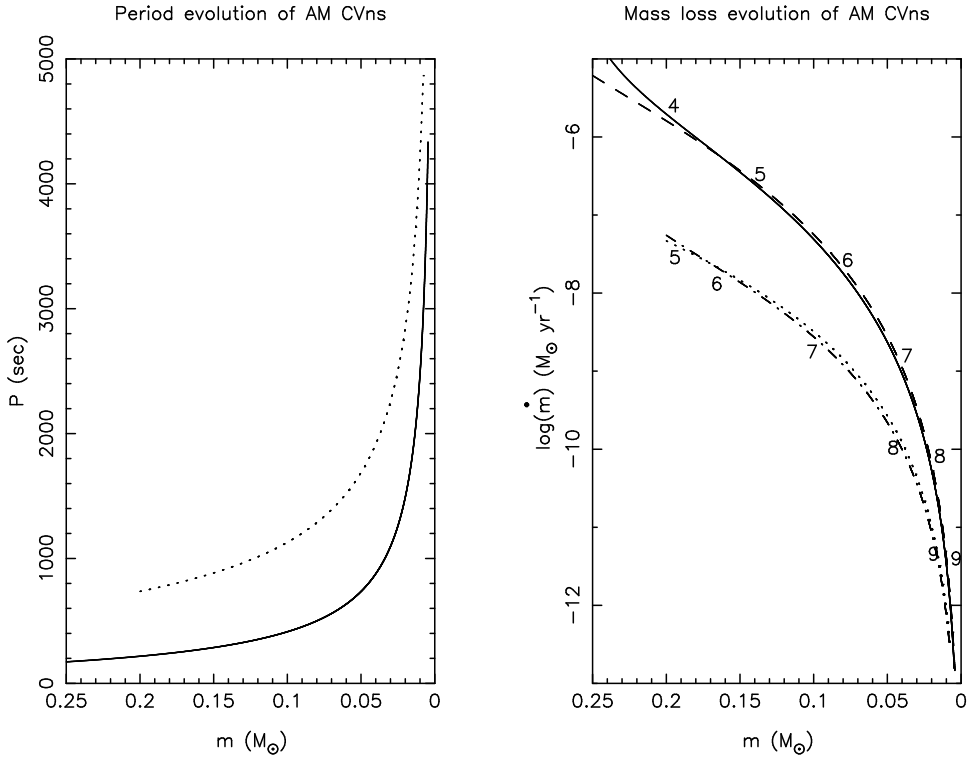}}
\caption[]{
  Examples of the evolution of AM CVn systems.  The left panel shows
  the evolution of the orbital period as function of the mass of the
  secondary (donor) star.  The right panel shows the change in the
  mass transfer rate during the evolution.  The solid and dashed lines
  are for white dwarf donor stars of initially 0.25\,\mbox{${\rm M}_{\sun}$}\ 
  transferring matter to a primary of initial mass of 0.4 and
  0.6\,\mbox{${\rm M}_{\sun}$}, respectively, assuming efficient coupling between the
  accretor spin and the orbital motion. The dash-dotted and dotted
  line are for a helium star donor, starting when the helium star
  becomes semi-degenerate (with a mass of 0.2\,\mbox{${\rm M}_{\sun}$}). Primaries are
  again 0.4 and 0.6\,\mbox{${\rm M}_{\sun}$}. The numbers along the lines indicate the
  logarithm of the time in years since the beginning of the mass
  transfer. 
}
\label{fig:PMdot}
\end{figure*}
 
Figure~\ref{fig:PMdot} presents two examples of the evolution of the
orbital period and the mass transfer rate for both families of AM CVn
systems. Initially the mass transfer rate is very high but within a
few million years it drops below $10^{-8}$ \mbox{${\rm M}_{\sun}$}\ yr$^{-1}$.  In the
same time interval the orbital period increases from a few minutes, in
the case of the white dwarf family, or from a little over 10 minutes,
for the helium star family, to a few thousand seconds. The
semi-degenerate donor systems have lower mass transfer rates and
larger periods for the same donor mass due to their larger radii.  The
fact that the period is independent of the accretor mass ($M$) is a
consequence of Eq.~(\ref{eq:r_L}) and Keplers 3rd law leading to $P
\propto (R^3/m)^{1/2}$.

\section{The population of AM CVn stars}\label{sec:popsynth}

We used the population synthesis program \textsf{SeBa}, as described
in detail in \citet{pv96}, \citet{py98} and \citet{nyp+00} to model
the progenitor populations. We follow model A of \citet{nyp+00}, which
has an IMF after \citet{ms79} and flat initial distributions over
the mass ratio of the components and the logarithm of the orbital
separation and a thermal eccentricity distribution. We assume an initial
binary fraction of 50\% and that the star formation decreases
exponentially with time, which is different from other studies of
close double white dwarfs that assume a constant star formation rate.
The mass transfer between a giant and a main-sequence star of
comparable mass is treated with an ``angular momentum formalism''
which does not result in a strong spiral-in \citep{nvy+00}.

\subsection{The total population}\label{results:total}

\begin{figure*}
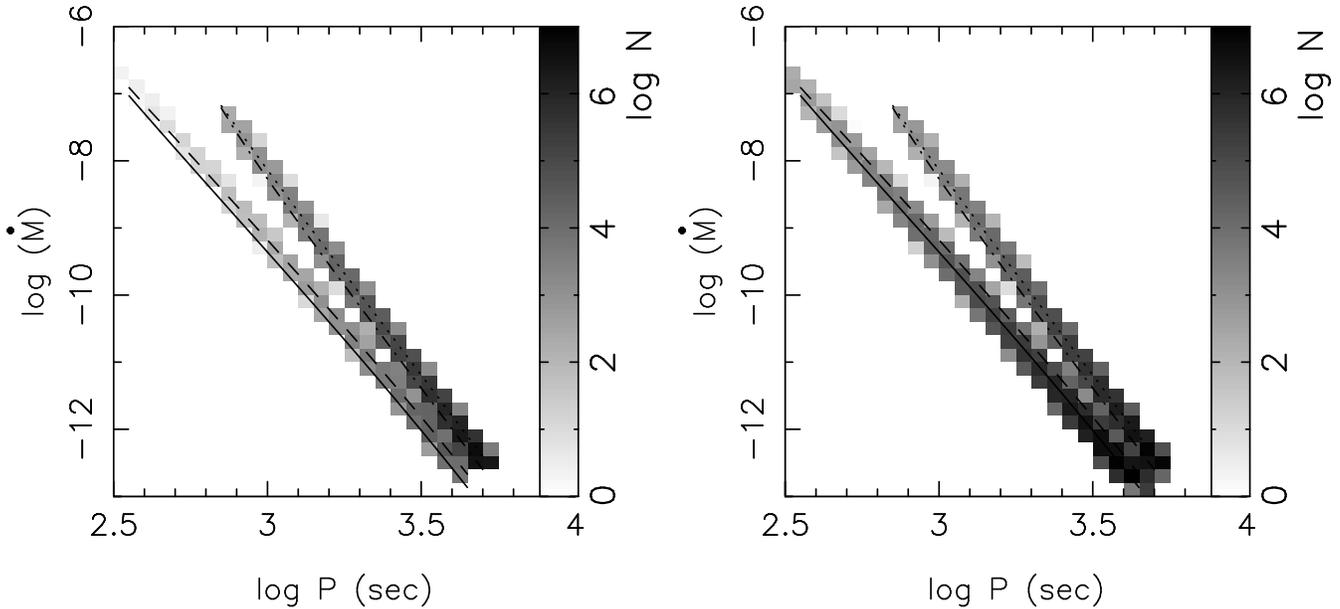

\begin{center}
\begin{minipage}{0.49\textwidth}
\psfig{figure=H2552f05.ps,width=\columnwidth,angle=-90,clip=} 
\end{minipage}
\begin{minipage}{0.49\textwidth}
\psfig{figure=H2552f06.ps,width=\columnwidth,angle=-90,clip=} 
\end{minipage}
\end{center}
\caption[]{The current Galactic population of AM CVn systems of the two
  families for models I (left) and II (right). The grey scale indicates the
  \emph{logarithm} of the number of systems. The upper branch is the
  helium star family; the lower branch the white dwarf family.
  The lines show the orbital period and mass transfer rate evolution and
  correspond to the lines in Fig.~\ref{fig:PMdot}.
}
\label{AMCVn_logPMdot}
\end{figure*}

We generate the population of close double white dwarfs and helium
stars with white dwarf companions and select the AM CVn star
progenitors according to the criteria for the formation of the AM CVn
stars as described above. We calculate the birthrate of AM CVn stars,
and evolve every system according to the recipe described in
Sect.~\ref{model}, to obtain the total number of systems currently
present in the Galaxy (Table~\ref{tab:birthrates}) and their
distribution over orbital periods and mass loss rates
(Fig.~\ref{AMCVn_logPMdot}).

The absence of an effective coupling between the accretor spin and the
orbital motion (model I) reduces the current birth rate AM CVn stars
from the white dwarf family by two orders of magnitude as compared to
the case of effective coupling (model II).  The fraction of close
double white dwarfs which fill their Roche lobes and continue their
evolution as AM CVn stars is 21\% in model II but only 0.2\% in model
I (see also Fig.~\ref{fig:MsMp}).
  
In model I the population of AM CVn stars is totally dominated by the
helium star family. In model II where tidal coupling is efficient both
families have a comparable contribution to the population.  Increasing
the mass of the critical layer for ELD from 0.15\,\mbox{${\rm M}_{\sun}$}\ to
0.3\,\mbox{${\rm M}_{\sun}$}\ almost doubles the current birth rate of the systems which
are able to enter the semi-degenerate branch of the evolution.  In the
latter case almost all helium star binaries that transfer matter to a
white dwarf in a stable way eventually become AM CVn systems (see
Fig.~\ref{hewd}).

In Fig.~\ref{AMCVn_logPMdot} we show the total current population of
AM CVn systems in the Galaxy in our model.  The evolutionary paths of
both families are indicated with the curves (see also
Fig.~\ref{fig:PMdot}). Table~\ref{tab:birthrates} gives the total
number of systems currently present in the Galaxy.  The evolution of
the systems decelerates with time and as a result the vast majority of
the systems has orbital periods larger than one hour. The evolutionary
tracks for the two families do not converge, since the mass loss of
the helium stars prevents their descendants from recovering thermal
equilibrium in the lifetime of the Galactic disk (see
sect.\ref{formation:hewd}).

The minimum donor mass attainable within the lifetime of the
Galactic disk is $\sim 0.005\,\mbox{${\rm M}_{\sun}$}$\ for the descendants of the
helium white dwarfs and $\sim 0.007\,\mbox{${\rm M}_{\sun}$}$\ for the descendants of
the helium stars. This is still far from the limit of $\sim
0.001\,\mbox{${\rm M}_{\sun}$}$\, where the electrostatic forces in their interiors will
start to dominate the gravitational force, the mass-radius relation
will become $R \propto M^{1/3}$\ \citep{zs69}, and the mass transfer
will cease.

In Table~\ref{tab:birthrates} we give the local space density of AM
CVn systems estimated from their total number and the Galactic
distribution of stars, for which we adopt 
\begin{equation}\label{eq:rho}
        \rho(R, z) = \rho_{\rm 0} \; e^{-R/H} \; 
                     \mbox{sech}(z/h)^2  \quad \mbox{pc}^{-3}
\end{equation}
as in \citep{nyp+00}. Here $H$ = 2.5 kpc \citep{sac97} and $h$ = 200
pc, neglecting the age and mass dependence of $h$.  These estimates
can not be compared directly to the space density, estimated from the
observations: $3 \, \times \, 10^{-6}$ pc$^{-3}$ \citep{war95}. In our
model the space density is dominated by the long-period, dim systems,
while Warner's estimate is based on the observed systems which are
relatively bright.  For a comparison of the observed and predicted
populations we have to consider selection effects.

\begin{table}[!t]
\caption[]{
Birth rate and number of AM CVn systems in the Galaxy.
The first column gives the model name (Sect.~\ref{sec:two_fam})
followed by the current Galactic birthrate ($\nu$ in yr$^{-1}$), the
total number of systems in the Galaxy (\#) and the number of observable
systems with V $<$ 15 (\#~obs).  The last column ($\sigma$ in
pc$^{-3}$) gives the local space density of AM CVn stars for each
model.  Due to selection effects the number of observable systems is
quite uncertain (see Sect.~\ref{selection}).  
}
\begin{tabular}{l|ccc|ccc|c}
\hline
 &   \multicolumn{3}{c|}{white dwarf family} &  \multicolumn{3}{c|}{He-star  family} &  \\
Mod.    & $\nu$       & \# & \#~obs & $\nu$ & \# & \#~obs & $\sigma$\\
    & $10^{-3}$ &  $10^7$  &     & 10$^{-3}$ & $10^7$  & & $10^{-4}$    \\ \hline
I   &  0.04     & 0.02    & ~~1 & 0.9       & 1.8     & 32 & 0.4\\  
II  &  4.7~~    &  4.9~~ & 54  & 1.6       & 3.1       & 62 & 1.7\\ \hline
\end{tabular}

\label{tab:birthrates}
\end{table}

\subsection{Observational selection effects: from the total population 
  to the observable population}\label{selection}

The known systems are typically discovered as faint blue stars (and
identified with DB white dwarfs), as high proper motion stars, or as
highly variable stars \citep[see for the history of detection of most
of these stars][]{ull94, war95}. The observed systems thus do not
have the statistical properties of a magnitude limited sample.

Moreover, the luminosity of AM CVn stars comes mainly from the disk in
most cases.  Despite the fact that several helium disk models are
available \citep[e.g.][]{sma83,can84,to97,ew00} there is no easy way
to estimate magnitude of the disk.  Therefore, we compute the visual
magnitude of the systems from very simple assumptions, to get a notion
of the effect of observational selection upon the sample of
interacting white dwarfs.

The luminosity provided by accretion is
\begin{equation}
L_{\rm acc} \approx 0.5 \; G \; M \dot{m} \; \left( \frac{1}{R} -
  \frac{1}{R_{\rm L1}} \right).
\end{equation}
Here $R$ is the radius of the accretor and $R_{\rm L1}$ is the
distance of the first Lagrangian point to the centre of mass of the
accretor. We use an ``average'' temperature of the disk \citep[see][]{wad84},
which may be then obtained from $L = S \sigma T^4$, where $S$ is the
total surface of the disk:
\begin{equation}
S =  2 \pi (R_{\rm out}^2 - R_{\rm WD}^2).
\end{equation}
We use $R_{\rm out} = 0.7 R_{\rm L1}$.  The visual magnitude of the
binary is then computed from the effective temperature and the
bolometric correction \citep[taken from][]{kui38}, assuming that the
disk is a black body . This allows us to construct a magnitude limited
sample by estimating the fraction of the Galactic volume in which any
system in our theoretical sample may be observed as it evolves.

\begin{table}
\caption[]{Orbital
  periods, visual magnitudes and theoretical mass estimates for known
  and candidate AM CVn stars.
}
\begin{center}
\label{obs}
\begin{tabular}{lccccl}
\hline 
 Name & Period      & $m_v$ &  m   &  m    & Ref. \\
      & ${\rm sec}$ &       & (ZS) &  (TF) &    \\
\hline 
AM CVn & 1028.7 & 14.1-14.2 &  0.033 &  0.114 &  1 \\
HP Lib & 1119~~ & 13.6 &  0.030&  0.099 &  2 \\
CR Boo & 1471.3 & 13.0-18.0 &  0.021 &  0.062 &  3 \\
V803 Cen & 1611~~ & 13.2-17.4 &  0.019 &  0.054 &  2 \\
CP Eri & 1724~~ & 16.5-19.7 &  0.017 &  0.048 &  2 \\
GP Com & 2970~~    & 15.7-16.0     &  0.008 &  0.019 &  2 \\
RX~J1914+24 & 569~ & $>$ 19.7 &  0.068 &      - & 4 \\
KL Dra          &      & 16.8-20      &  &   &  5     \\
\hline
\end{tabular}\\
\end{center}
Theoretical mass estimates (in \mbox{${\rm M}_{\sun}$}) obtained from Eq.~(\ref{eq:deg})
are labelled by ZS, estimates from Eq.~(\ref{eq:semi}) by TF.\\
References: (1) \citet{phs93}, (2) \citet{war95}, (3)\citet{pwn+97},
(4) \citet{chm+98}, (5) \citet{sch98}
\end{table}

\begin{figure*}
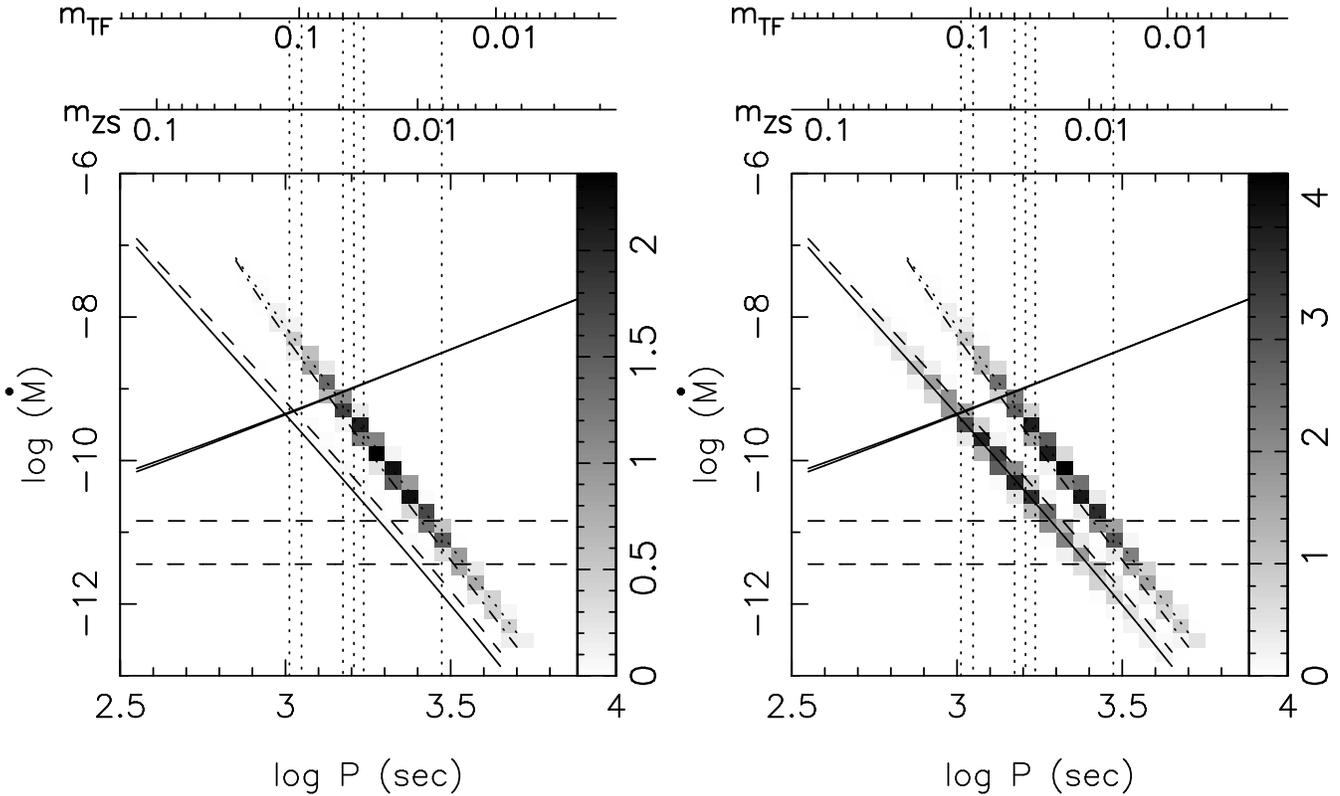

\begin{minipage}{0.49\textwidth}
\psfig{figure=H2552f07.ps,width=\columnwidth,angle=-90} 
\end{minipage}
\begin{minipage}{0.49\textwidth}
\psfig{figure=H2552f08.ps,width=\columnwidth,angle=-90} 
\end{minipage}
\caption[]{
  Magnitude limited sample ($V_{\rm lim} = 15$) of the theoretical
  population of AM CVn stars for model I (left) and model II (right).
  The grey scale gives the number of systems, like in
  Fig.~\ref{AMCVn_logPMdot} but now on a linear scale (upper branch
  for the helium star family; lower branch for the white dwarf
  family). The selection criteria are described in
  Sect.~\ref{selection}. The periods of the observed systems
  (Table~\ref{obs}) are indicated with the vertical dotted lines. The
  stability limits for the helium accretion disk according to
  \citet{to97} are plotted as the solid slanted and dashed
  horizontal lines.  Between these lines the disk is expected to be
  unstable. The upper dashed line is for an accretor mass of 0.5
  \mbox{${\rm M}_{\sun}$}, the lower for a 1.0 \mbox{${\rm M}_{\sun}$}\ accretor.  The rulers at the top
  indicate the theoretical relation between the period and the mass
  for the mass-radius relations of the two AM CVn star families given
  by Eqs.~(\ref{eq:deg}) and (\ref{eq:semi}). }
\label{AMCVn_PMdot}
\end{figure*}                                                                           
We derive $P - \dot{M}$ diagrams for both models, similar to the ones
for the total population, but now for the ``observable'' population,
which we limit by $V = 15$. Changing $V_{\rm lim}$ doesn't change the
character of graphs, since only the nearby systems are visible.  The
expected number of observable systems for the two families of
progenitors is given in Tab.~\ref{tab:birthrates} and shown in
Fig.~\ref{AMCVn_PMdot}. The observable sample comprises only one star
for every million AM CVn stars that exists in the Galaxy. A large
number of AM CVn stars may be found among very faint white dwarfs
which are expected to be of the non-DA variety due to the fact, that
the accreted material is helium or a carbon-oxygen mixture.

In the ``inefficient'' model I about one in 30 observed systems is
from the white dwarf family. This is a considerably higher fraction
than in the total AM CVn population where it is only one out of 100
systems.  In the ``efficient'' model II, the white dwarf family
comprises $\sim 60\%$\ of the total population and $\sim 50\%$\ of the
``observable'' one. The ratio of the total number of systems of the
white dwarf family in models I and II is not proportional to the ratio
of their current birthrates.  This reflects the star formation history
and the fact that the progenitors of the donors in model I are low
mass stars that live long before they form a white dwarf.  In model I
the fraction of the observable systems which belong to the white dwarf
family is higher than the fraction of the total number of systems that
belong to this family. This is caused by the fact that the accretors
in these systems are more massive (see Fig.~\ref{fig:MsMp}), thus
smaller, giving rise to higher accretion luminosities.

To compare our model with the observations, we list the orbital
periods and the observed magnitude ranges for the known and candidate
AM CVn stars in Table~\ref{obs}. For AM CVn we give $P_{\rm orb}$ as
inferred by \citet{phs93} and confirmed as a result of a large
photometry campaign \citep{spk+99} and a spectroscopic study
\citep{nsp00}. For the remaining systems we follow the original
determinations or \citet{war95}. Most AM CVn stars show multiple
periods, but these are close together and do not influence our
qualitative analysis. KL Dra is identified as an AM CVn type star by
its spectrum \citep{jgc+98}, but still awaits determination of its
period. The periods of the observed AM CVn stars are shown in
Fig.~\ref{AMCVn_PMdot} as the vertical dotted lines. The period of
RX~J1914+24 is not plotted because this system was discovered as an
X-ray source and it is optically much fainter than the limit used
here.

Figures~\ref{AMCVn_logPMdot} and \ref{AMCVn_PMdot} show that the
uncertainty in both models and observational selection effects make
it hard to argue which systems belong to which family. According to
model I the descendants of close double white dwarfs are very rare.
However, in that case one might not expect two observed systems at
short periods (AM CVn and HP Lib). In both models I and II, systems
with long periods (like GP Com) are more likely to descend from the
helium star family. In the spectrum of GP Com, however, \citet{mhr91}
found evidence for hydrogen burning ashes in the disk, but no traces
of helium burning, viz. very low carbon and oxygen abundances.  It is
not likely that any progenitor of the helium star family completely
skipped helium burning. More probably, this system belongs to the
white dwarf family.

Most systems in the ``observable'' model population have orbital
periods similar to the periods of the observed AM CVn stars that show
large brightness variations; thus most modelled systems are expected
to be variable. These brightness variations have been interpreted as
a result of a thermal instability of helium disks \citep{sma83}. In
Fig.~\ref{AMCVn_PMdot} we show the thermal stability limits for helium
accretion disks as derived by \citet{to97}: above the solid line the
disks are expected to be hot and stable; below the horizontal dashed
lines the disks are cool and stable and in between the disks are
unstable. Note that the vast majority of the total Galactic model
population (Fig.~\ref{AMCVn_logPMdot}) is expected to have cool stable
disks according to the thermal instability model, preventing them from
being detected by their variability.

The period distributions of the ``observable'' population in our
models agree quite well with the observed population of AM CVn
stars. Better modelling of the selection effects is, however,
necessary.

\subsection{Individual systems}

Table~\ref{obs} gives theoretical estimates of the masses of the donor
stars in the observed AM CVn stars, derived from the relation between
the orbital period and the mass of the donor (see
Sect.~\ref{sec:two_fam} and Fig.~\ref{AMCVn_PMdot}).

AM CVn stars may be subject to tidal instability due to which the disk
becomes eccentric and starts precessing. Such instabilities are used
to explain the superhump phenomenon in dwarf novae \citep{whi88}.

For AM CVn and CR Boo the observed 1051.2s \citep{spb+98} and 1492.8s
\citep{pwn+97} periodicities are interpreted as superhump periods.
Following \citet{war95} we compute the mass ratio of the binary system
using the orbital period ($P_{\rm orb}$) and the superhump period
($P_{\rm s}$) via:
\begin{equation}\label{eq:pops}
        \frac {P_{\rm s}}{P_{\rm s} - P_{\rm orb}} 
        \approx 
        3.73 \, \frac{1+q}{q}.
\end{equation} 
This results is $q$ = 0.087 and 0.057 for AM CVn and CR Boo
respectively. Assuming that they belong to the white dwarf family
their accretor masses are $M=0.38\,\mbox{${\rm M}_{\sun}$}$\ and $M=0.37 \,\mbox{${\rm M}_{\sun}$}$.
These values are at the lower end of the predicted distribution. If we
apply the semi-degenerate mass - radius relation, the estimated masses
of the accretors are high, even close to the Chandrasekhar
  mass for AM CV. The formation of systems with high-mass accretors
  has a low probability (see Fig.~\ref{hewd}), which suggests that
either Eq.~(\ref{eq:pops}) is not applicable for helium disks or
alternatively that these binaries do not belong to the helium star
family.

Maybe the most intriguing system is RX~J1914.4+245; detected by {\it
  ROSAT} \citep{mhg+96} and classified as an intermediate polar,
because its X-ray flux is modulated with a 569 s period, typical for
the spin periods of the white dwarfs in intermediate polars.
\citet{chm+98} and \citet{rcw+00} suggest that it is a double
degenerate polar with an orbital period equal to the spin period of
the accreting white dwarf. The mass transfer rate in this system,
inferred from its period ($\dot{m} \approx 1.8 \, \times \, 10^{-8} \,
\mbox{${\rm M}_{\sun}$}$ yr$^{-1}$) is consistent with the value deduced from the {\it
  ROSAT} PSPC data \citep{chm+98} if the distance is $\sim$100 pc.

Even though polars have no disk, the coupling between the accretor and
donor is efficient due to the strong magnetic field of the accretor.
We therefore anticipate that Eq.~(\ref{eq:q_II}) applies without the
correction introduced by Eq.~(\ref{eq:q_I}).  It may well be that
magnetic systems in which the coupling is maintained by a magnetic
field form the majority of stable AM CVn systems of the white dwarf
family. We do not expect this system to belong to the helium star
family, since its period is below the typical period minimum for the
majority of the binaries in this family.

RX~J0439.8-809 may be a Large Magellanic Cloud relative of the
Galactic AM CVn systems. This system was also first detected by {\it
  ROSAT} \citep{grei94}. Available X-ray, UV- and optical data
suggest, that the binary may consist of two degenerate stars and have
an orbital period $< 35$ min \citep{trh+97,tgb99}.
 
RX~J1914.4+245 and RX~J0439.8-809 show that it is possible to
detect optically faint AM CVn stars in supersoft X-rays, especially in
other galaxies. The possibility of supersoft X-rays emission by AM CVn
stars was discussed by \citet{ty96}.  There are two probable sources
for the emission: the accreted helium may burn stationary at the
surface of the white dwarf if $\dot{m} \sim 10^{-6} \, \mbox{${\rm M}_{\sun}$}$
yr$^{-1}$ and/or the accretion disk may be sufficiently hot in the
same range of accretion rates. However, the required high accretion
rate makes such supersolf X-ray sources short-living (see
Fig.~\ref{fig:PMdot}) and, therefore, not numerous. Note that AM CVn,
CR Boo, V803 Cen, CP Eri and GP Com are also weak X-ray sources
\citep{ull95a}.

The most recently found suspected AM CVn star, KL~Dra, is also
variable. Therefore we expect it to lie in the same period range as CR
Boo, V803 Cen and CP Eri. Taking the limits for stability as given by
\citet{to97} we expect the orbital period to be between 20 and 50
minutes (Fig.~\ref{AMCVn_PMdot}).

\section{Discussion}\label{discussion}

A population synthesis study for AM CVn stars (and related systems)
was done by \citet{ty96}, who considered only an ``efficient'' model.
Their derived birthrate for the white dwarf family is $1.3 \times
10^{-2}$ yr$^{-1}$, a factor three higher than the value in our
``efficient'' model.  This difference can in part be explained by the
different treatment of the mass transfer from a giant to a main
sequence star of comparable mass \citep[see][]{nvy+00,nyp+00}.  Most
close double white dwarfs in our model have a mass ratio close to
unity for which stable mass transfer is impossible
(Sect.~\ref{model}), while in the model of \citet{ty96} they
predominantly have $q \sim 0.5 - 0.7$ which is more favourable for
stable mass transfer.  Our higher integrated star formation rate only
partly compensates for the loss of stable systems.

Another difference is that \citet{ty96} conclude that the helium star
family (non-degenerate helium stars in their terminology) do not
contribute significantly to the AM CVn population. This is a
consequence of their assumption that these systems, after the period
minimum, live only for $10^8$\,yr. In contrast, our calculations show
that their evolution is limited only by the lifetime of the Galactic
disk.  Tutukov \& Yungelson estimate the total number of AM CVn stars from
the helium star family as $(1.9 - 4.6) \times 10^{5}$ depending on the
assumptions about the consequences of the accretion of helium. We find
$2 \times 10^{7}$ even when we let ELD destroy the systems which
accrete only 0.15 \mbox{${\rm M}_{\sun}$}.

An additional complication is the possibility of the formation of a
common envelope for systems where the accretion rate exceeds the rate
of stationary helium burning at the surface of the accreting white
dwarf ($\sim 10^{-6}\,\mbox{${\rm M}_{\sun}$}$ yr$^{-1}$). If such a common envelope
forms the components of system may well merge. If it happens, it will
occur directly after the Roche-lobe contact, when the highest
accretion rate occurs. We do not consider this possibility in our
model, because it involves too many additional (and unknown)
parameters.  Applying only the requirement that stable systems should
accrete below the Eddington rate, we may overestimate the birthrate of
AM CVn stars.

We did not discuss the Roche lobe overflow by low mass stars with
almost exhausted hydrogen cores ($X_c \sim 0.01$) which may also
result in the formation of helium transferring systems with orbital
periods $\sim 10$\,min \citep{tfe+87} because of its extremely low
probability.

The prescription for ELD is related to the problem of SN\,Ia
progenitors. In model I almost all accretors in the helium star family
with initial $M \geq 0.6$\,\mbox{${\rm M}_{\sun}$}\ ``explode'' and the ELD rate is
close to 0.001 yr$^{-1}$.  If ELDs really produce SNe\,Ia, they may
contribute about 25\% of their currently inferred Galactic rate. In
model II 0.3\,\mbox{${\rm M}_{\sun}$}\ must be accreted prior to the explosion, and the
ELD rate is only about $4 \, \times \,10^{-4}$ yr$^{-1}$. Even in
model I we find a much lower ELD rate than \citet[who find
0.005]{ty96}. This is partly due to a lower birthrate, but also to the
different treatment of the mass transfer from a giant to a main
sequence star of comparable mass \citep{nvy+00}, which causes the
accretors in our model mainly to have masses below 0.6 \mbox{${\rm M}_{\sun}$}. Such
systems probably never experience ELD.

In model II the accretors in both families may accrete so much matter
that they reach the Chandrasekhar mass. The rates for the white dwarf
and helium star families for this process are $3 \times 10^{-6}$ and
$5 \times 10^{-5}$ yr$^{-1}$.

In both families the accretors can be helium white dwarfs (see
Figs.~\ref{fig:MsMp} and \ref{hewd}).  It was shown by \citet{ns77}
that accretion of helium onto helium white dwarfs with $\dot{m} = (1 -
4) \, \times \, 10^{-8} \, \mbox{${\rm M}_{\sun}$}$ yr$^{-1}$\ results either in a
helium shell flash (at the upper limit of the accretion rates) or in
central detonation which disrupts the white dwarf (for lower
$\dot{m}$). The detonation occurs only when the mass of the accretor
grows to $\sim 0.7\,\mbox{${\rm M}_{\sun}$}$.  In our calculations this happens for the
helium star family at a rate of $\sim 4 \, \times \, 10^{-6}$
yr$^{-1}$. For the white dwarf family it happens only in model II, at
a rate of $\sim 2 \, \times \, 10^{-6}$ yr$^{-1}$.

\section{Conclusions}\label{conclusions}

We study the formation of AM CVn stars from (i) close detached double
white dwarfs which become semi-detached and (ii) helium stars that
transfer matter to a white dwarf and stop burning helium due to mass
loss and become dim and semi-degenerate.

We find that, with our assumptions, in all cases where a double white
dwarf potentially can form an AM CVn star no accretion disk will be
formed in the initial phase of mass transfer. Normally the disk
provides the feedback of angular momentum to the orbit, stabilising
the mass transfer.  In absence of a disk, the stability of the mass
transfer in the semi-detached white dwarf binary depends critically on
the efficiency of the coupling between the accretor and the donor.  If
this coupling is not efficient most systems merge, and the formation
rate of AM CVn stars from double white dwarfs becomes very low.  In
this case it is possible that magnetically coupled systems are
  almost the only ones to survive. RX~J1914.4+245 may be such a
system.

In the second channel the formation of AM CVn stars may be
prevented by explosive burning of the accumulated helium layer which
may cause detonation of the CO white dwarf accretor and the disruption
of the system.

We combine our population synthesis results into two models, an
``efficient'' model in which the stability of mass transfer is not
affected by the absence of an accretion disk and the explosive helium
burning disrupting the system happens when 0.3 \mbox{${\rm M}_{\sun}$}\ is accumulated
and an ``inefficient'' model in which the absence of an accretion disk
is very important and the explosive helium disrupting the system
happens already when 0.15 \mbox{${\rm M}_{\sun}$}\ is accumulated. Applying very simple
selection effects we estimate that in the ``inefficient'' model only
one in 30 potentially observed systems descends from double white
dwarfs. In the ``efficient'' model both families produce comparable
numbers of observable systems. The observed systems fall roughly in
the expected range of periods for a magnitude limited sample. 

We conclude that to learn more about the AM CVn population both theory
(stability of the mass transfer and helium accretion disks) and
observations (especially the distances and the completeness of the
sample) need to be improved.

\begin{acknowledgements}
  LRY and SPZ acknowledge the warm hospitality of the Astronomical
  Institute ``Anton Pannekoek''. This work was supported by NWO
  Spinoza grant 08-0 to E.~P.~J.~van den Heuvel, RFBR grant
  99-02-16037, the ``Astronomy and Space Research Program'' (project
  1.4.4.1) and by NASA through Hubble Fellowship grant HF-01112.01-98A
  awarded (to SPZ) by the Space Telescope Science Institute, which is
  operated by the Association of Universities for Research in
  Astronomy, Inc., for NASA under contract NAS\,5-26555.
\end{acknowledgements}

\bibliography{journals,binaries}
\bibliographystyle{apj}
\end{document}